\documentclass[%
aip,
sd,%
amsmath,amssymb,
author-numerical,%
]{revtex4-1}

\usepackage{graphicx}% Include figure files
\usepackage{dcolumn}% Align table columns on decimal point
\usepackage{bm}% bold math
\usepackage[dvipsnames]{xcolor}
\usepackage[normalem]{ulem}
\usepackage[title]{appendix}
\usepackage{comment}
\usepackage{float}

\graphicspath{{./figures/}}
\begin{document}

\title{ Reservoir Computers Modal Decomposition and Optimization}
%\date{October 2020}

	\author{Chad Nathe}
	\affiliation{ 
		Mechanical Engineering Department, University of New Mexico, Albuquerque, NM 87131}

\author{Enrico Del Frate}
	\affiliation{ 
		Mechanical Engineering Department, University of New Mexico, Albuquerque, NM 87131}

\author{Thomas Carroll}
	\affiliation{ 
		U.S. Naval Research Laboratory, Washington, DC 20375, USA}

\author{Louis Pecora}
	\affiliation{ 
		U.S. Naval Research Laboratory, Washington, DC 20375, USA}

\author{Afroza Shirin}
	\affiliation{ 
		Mechanical Engineering Department, University of New Mexico, Albuquerque, NM 87131}
		
	\author{Francesco Sorrentino}
	\email{fsorrent@unm.edu}
	\affiliation{
		Mechanical Engineering Department, University of New Mexico, Albuquerque, NM 87131}

\begin{abstract}
    The topology of a network associated with a reservoir computer is often taken so that the connectivity and the weights are chosen randomly. Optimization is hardly considered as the parameter space is typically too large. Here we investigate this problem for a class of reservoir computers for which we obtain a decomposition of the reservoir dynamics into modes, which can be computed independently of one another.  Each mode depends on an eigenvalue of the network adjacency matrix. We then take a parametric approach in which the eigenvalues are parameters that can be appropriately designed and optimized. In addition, we introduce the application of a time shift to each individual mode. We show that manipulations of the individual modes, either in terms of the eigenvalues or the time shifts, can lead to dramatic reductions in the training error. 
\end{abstract}

\maketitle

%\section{Introduction}

A reservoir computer (RC) is a complex nonlinear dynamical system that is used for processing and analyzing empirical data, see e.g. \cite{jaeger2001echo, schrauwen2007overview, natschlager2002liquid, maass2002real, martinenghi2012photonic,brunner2013parallel,nakajima2015information, hermans2015photonic,vinckier2015high,duport2016fully,larger2017high}, 
modeling of complex dynamical systems \cite{suykens2012artificial}, 
speech recognition \cite{crutchfield2010introduction},  learning of context free and context sensitive languages \cite{rodriguez2001simple, gers2001lstm},
the reconstruction and prediction of chaotic attractors \cite{lu2018attractor, zimmermann2018observing,antonik2018using, jaeger2004harnessing,pathak2017using,pathak2018model},
image recognition \cite{jalalvand2018application}, and
control of robotic systems \cite{graves2004biologically, robinson1994application,lukovsevivcius2012reservoir}. 
A typical RC consists of a set of nodes coupled together to form a network. Each node of the RC evolves in time in response to an input signal that is fed into the reservoir. An output signal is then generated from the time evolutions of the RC nodes.
In a RC, the output connections (those that connect the RC nodes to the output) are trained to produce a best fit between the output signal and a training signal related to the original input signal. On the other hand, the connections between the nodes of the reservoir are constant parameters of the system. As a result, RCs are easier to analyze than other machine learning tools for which all the connections are typically  trained.

{Reference \cite{carroll2019network} studied the effects of the network topology on the performance of a reservoir computer and focused on the sparsity of the connections and the presence of network symmetries. Recent work  has analyzed linear reservoir computers \cite{boyd1985fading, bollt2020explaining} and pointed out a connection with the theory of dynamic mode decomposition \cite{schmid2010dynamic}. A common assumption is that nonlinear reservoirs can outperform linear reservoirs \cite{bollt2020explaining}. Optimizing the hyperparameters of a reservoir computer is often done, but optimizing the connections between the RC nodes is more difficult due to the high-dimensional parameter space. The standard recipe is to use random matrices. Our analysis that follows shows that under certain conditions, the reservoir equations can be rewritten in an equivalent form which corresponds to individual uncoupled nodes, which are easier to optimize.  }

%\section{Reservoir Dynamics}
We consider a reservoir computer modeled by the following dynamical equations in continuous time,

\begin{equation} \label{eq:1}
    \dot r_i(t) = F\Bigl( r_i(t), \sum_{j=1}^N A_{ij}r_j(t), s_1(t), s_2(t),..., s_l(t) \Bigr), \quad i=1,...,N,
\end{equation}
where $r_i$ is the scalar state of node $i$ of the reservoir, $N$ is the number of nodes, the adjacency matrix $A=\{A_{ij}\}$ describes the connectivity between the network nodes, and $s_1,s_2,...s_l$ are input signals to the reservoir.  These can represent different data being fed into the reservoir, such as in a weather prediction application, a rainfall time series, a wind time series, a humidity time series, and so on.  These input signals are in general a function of an underlying process to which the reservoir is applied. The  training signal $g(t)$ is another signal from the same underlying process which is related to the input signals through a complex relation (e.g., in the weather prediction application, a temperature time series.) The function $F:R^{l+2} \rightarrow R$ determines the particular dynamics of the reservoir nodes. 
Next we focus on a specific class of reservoirs, {which possess the property of universality \cite{grigoryeva2018universal}},  described by the following set of equations,
\begin{equation} \label{eq:s}
    \dot r_i(t) = \alpha \Bigl(1+\epsilon s_2(t)\Bigr) r_i(t)+ \sum_{j=1}^N A_{ij}r_j(t) + w_i s_1(t), \quad i=1,...,N,
\end{equation}
where $l=2$ and $s_1$ and $s_2$ are the two input signals. In what follows, we will often refer to Eq.\ \eqref{eq:s} as that of a linear reservoir computer. The particular process to generate the adjacency matrix $A$ is described in the Supplementary Information.
 %The off-diagonal entries of the adjacency matrix $A$, $A_{ij}=A_{ji}$, $j \neq i$, are integers uniformly randomly drawn from the set $\{-1,0,1\}$.
In the rest of this paper we set $N=100$ and we also assume for simplicity that $A$ is symmetric, $A=A^T$.
The coefficients $w_i$ represent the weight by which the input signal is multiplied in the dynamics of node $i$. These are also typically randomly chosen \cite{carroll2019network}.

The underlying process we want to model may evolve in time based on a set of deterministic (chaotic) equations, such as the equations of the Lorenz chaotic attractor, in the variables $x_L(t), y_L(t), z_L(t)$ (see Supplementary Information.)
One task we can give the reservoir is to reconstruct the $z_L(t)$ time evolution (training signal) from knowledge of either $x_L(t)$ or $y_L(t)$ or both (input signals). We will also consider other tasks for which the time series are generated by other  chaotic or periodic systems, such as the Hindmarsh-Rose system or the Duffing system (see Supplementary Information.)

In order to examine the accuracy of the reservoir computer relative to the dynamical system it is modeling, we must have a way to quantify how well the reservoir is able to reproduce the training signal $g(t)$ from knowledge of the input signals $s_1(t)$ and  $s_2(t)$. After integrating the reservoir equations for a long enough time, its dynamics can be described by the $ T \times (N+1)$ matrix,
\begin{equation}
\Omega =
\left[\begin{array}{ccccc}
    r_1(1) & r_2(1) & ... & r_N(1) & 1 \\
    r_2(1) & r_2(2) & ... & r_N(2) & 1\\
    \vdots & \vdots & \vdots &  \vdots  & \vdots\\
    r_1(T) & r_2(T) & ... & r_N(T) & 1\\
\end{array}\right]
\end{equation}
Here, $N$ is the number of nodes in the reservoir computer and $T$ is the amount of time-steps taken. We add a  column whose entries are all ones to account for any constant offset. The fit $\mathbf h = [h(1),h(2),...,h(T)]$ to the training signal  $\mathbf g = [g(1),g(2),...,g(T)]$ is computed as  $h(t)=\sum_{i=1}^N \kappa_i r_i(t)+\kappa_{N+1}$ (or, equivalently, in vectorial form
$\mathbf h = \Omega \pmb{\kappa}$),
where  the vector $\pmb{\kappa} = [\kappa_1,\kappa_2,...,\kappa_{N+1}]$, which contains a set of unknown coefficients to be determined. We set
\begin{equation} \label{kappa}
    \pmb{\kappa}= \Omega^\dagger {\mathbf{g}},
\end{equation}
where with the symbol $\Omega^\dagger$ we indicate the pseudo-inverse of the matrix $\Omega$.

From this, we can compute the training error \cite{doi:10.1063/1.5123733}$\Delta = \frac{\langle \Omega \pmb{\kappa} - \mathbf g \rangle}{\langle \mathbf g \rangle}$,
where the notation $\langle \rangle$ is defined 
$\langle \mathbf X \rangle = \sqrt{\frac{1}{T} \sum_{i=1}^T (X(i) - \mu)^2}$
for $\mathbf X$ any $T$-dimensional vector and $\mu = \frac{1}{T} \sum_{i=1}^T X(i)$.

\begin{figure}[H]
\centering
    \includegraphics[width= 0.9\textwidth]{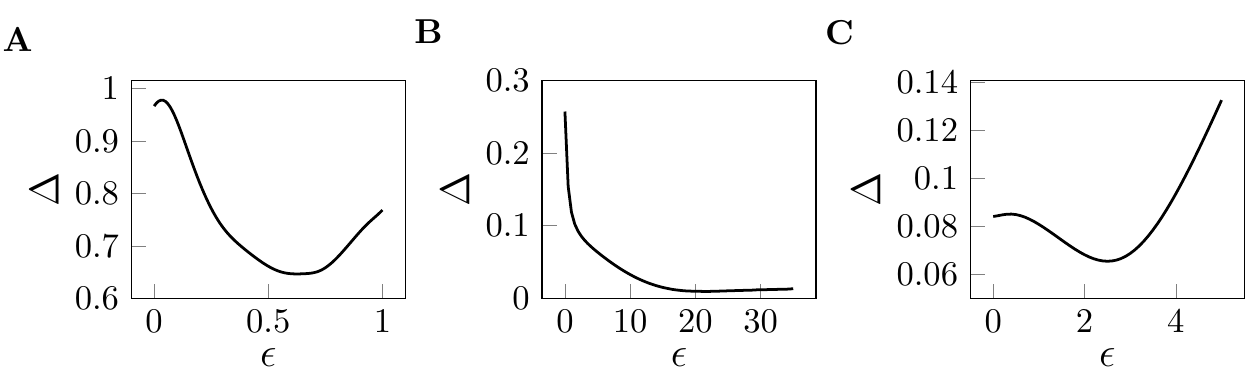}
    \caption{Plots of the training error vs. $\epsilon$. Here we plot the training error $\Delta$ for the following tasks in continuous time: Lorenz attractor (A),  Hindmarsh-Rose attractor (B), and the Duffing attractor (C). {In A and B $s_1(t)=s_2(t)=x(t)$ and $g(t)=z(t)$. In C $s_1(t)=s_2(t)=x(t)$ and $g(t)=y(t)$.} }
    \label{Continous vs Discrete 3tasks}
\end{figure}

Feeding the reservoir with more than one input signal or even driving the reservoir in different ways with the same input signal can lead to improved performance.
To see this, we perform numerical simulations 
in order to compare the single input with the $l=2$ input case of Eqs.\ \eqref{eq:s}. 
In Fig.\ \ref{Continous vs Discrete 3tasks} we plot the training error $\Delta$ vs the coefficient $\epsilon$ seen in Eq.\ \eqref{eq:s} (the case $\epsilon=0$ corresponds to no effect of $s_2(t)$ on the reservoir dynamics.) In this figure we deal with three different tasks, i.e., {reconstructing $g(t)=z(t)$ from $s_1(t)=s_2(t)=x(t)$ for the Lorenz chaotic system (A) and the Hindmarsh-Rose chaotic system (B), and reconstructing $g(t)=y(t)$ from $s_1(t)=s_2(t)=x(t)$ for the {Duffing periodic} system (C).} 

We see from these plots that as we increase $\epsilon$ the training error is first reduced and then it increases,  indicating the advantage of picking specific values of $\epsilon$. We have observed this type of relationship between the training error and $\epsilon$ in a large variety of situations, including discrete time reservoirs (see Supplementary Information). 
Our results show that typically two input reservoir computers \eqref{eq:s} are advantageous compared to the single input case ($\epsilon=0$).

We also considered the case that the reservoir is only driven by $s_2(t)$ and not by $s_1(t)$, i.e., for which the coefficients $w_i=0$. However, we found that for this case the training error was always close to $1$, which seems to indicate the advantage of the reservoir \eqref{eq:s} is limited to the case that both $s_1(t)$ and $s_2(t)$ are used.

%\section{Modal Decomposition of the Reservoir Dynamics}
{Next we obtain a modal decomposition for the reservoir dynamics \eqref{eq:s}.}
Our derivations that follow are obtained for continuous time, but analogous derivations can be obtained for discrete time (see Supplementary Information.)
 We first rewrite Eq. \eqref{eq:s} in vector form,

\begin{equation} \label{eq:rvector}
     \mathbf{\dot r}(t) = p_1(t) I \mathbf r(t) + A \mathbf r(t) + \mathbf w s_1(t),
 \end{equation}
where $I$ is the identity matrix, $p_1(t)=\alpha(1+\epsilon s_2(t))$ and,
\begin{equation}
    \mathbf{r}(t) =
    \left[\begin{array}{c}
    r_1(t)\\
    r_2(t)\\
    \vdots\\
    r_N(t)
    \end{array}\right]
    \;\;\;
    \mathbf{w} =
    \left[\begin{array}{c}
    w_1\\
    w_2\\
    \vdots\\
    w_N
    \end{array}\right]
\end{equation}

As we have set the adjacency matrix $A$ to be symmetric, it is also diagonalizable,
$A = V\Lambda V^T$
where $V$ is the matrix whose columns are the eigenvectors of $A$ and $\Lambda$ is a  diagonal matrix with the real eigenvalues of the matrix $A$ on the main diagonal. We pre-multiply Eq.\ \eqref{eq:rvector} by $V^T$ and after setting $\mathbf{q}(t) = V^T\mathbf r(t)$ and $\mathbf{c}(t) = V^T\mathbf w(t)$, we obtain, 
\begin{equation}
    \mathbf{\dot q}(t) = p_1(t) \mathbf{q}(t) + \Lambda \mathbf{q}(t) + \mathbf{c} s_1(t),
\end{equation}
which breaks up into a set of $N$ independent \textit{modes}, 
\begin{equation}\label{individual node reservior}
    {\dot q}_i(t) = p_1(t)q_i (t) + \lambda_i q_i (t) + c_i s_1(t), \quad i=1,...,N,
\end{equation}
with solution,
\begin{equation} \label{q}
    q_i(t)=\exp\Bigl(\int_0^t (\lambda_i+p_1(\tau)) d \tau \Bigr)  q_i(0)+ c_i \int_0^t \exp\Bigl(\int_\tau^t (\lambda_i+p_1(\rho)) d \rho \Bigr)  s(\tau) d \tau,
\end{equation}
where the first term on the left hand side of \eqref{q} is the free evolution which by the assumption of stability goes to zero for large $t$. For large $t$, each mode $q_i$ differs from the others through the coefficient $\lambda_i$, while the particular modal amplitude is given by $c_i$. We set
%=\sum_{i=1}^N \kappa_i \sum_{j=1}^N V_{ij} q_j(t)+\kappa_{N+1}=$
\begin{equation} \label{bestfit}
h(t) =\sum_j \kappa'_{j} q_j(t) + \kappa_{N+1},
\end{equation}
where 
\begin{equation} \label{NN}
\kappa'_j=\sum_{i=1}^N V_{ij} \kappa_i,
\end{equation}
i.e., $h(t)$ can be written as a linear combination of the modes $q_j(t)$. {It is important to emphasize that for large enough $t$ the particular value of $c_i$ becomes irrelevant in order to determine the best fit to the training signal, as in Eq.\ \eqref{bestfit} each mode is `rescaled' by a particular coefficient $\kappa'_i$. Another key observation is that the magnitude of $\int_0^t \exp\Bigl(\int_\tau^t (\lambda_i+p_1(\rho)) d \rho \Bigr)  s(\tau) d \tau$ will  depend on the value of $\lambda_i$. In practice, it may be convenient to properly rescale each mode to be
\begin{equation}
    \tilde{q}_i(t)=(\lambda_i+\bar{p}_1)^{-1} \int_0^t \exp\Bigl(\int_\tau^t (\lambda_i+p_1(\rho)) d \rho \Bigr)  s(\tau) d \tau,
\end{equation}
where $\bar{p}_1$ is a time-average value for $p_1$.}

We can now formulate the problem of finding the best fit $h(t)$ to the training signal 
in terms of the modes $q_1(t), q_2(t),...,q_N(t)$. 
To this end, we introduce  the $ T \times (N+1)$ matrix,
\begin{equation}
\Omega' =
\left[\begin{array}{ccccc}
    q_1(1) & q_2(1) & ... & q_N(1) & 1 \\
    q_1(2) & q_2(2) & ... & q_N(2) & 1\\
    \vdots & \vdots & \vdots &  \vdots  & \vdots\\
    q_1(T) & q_2(T) & ... & q_N(T) & 1\\
\end{array}\right]
\end{equation}
and we define the best fit to the training signal,
$
    \mathbf h' = \Omega' \pmb{\kappa}'$,
in the  vector $\pmb{\kappa}' = [\kappa'_1,\kappa'_2,...,\kappa'_{N+1}]$ contains a set of unknown coefficients to be determined. The best fit is obtained by setting
$
    \pmb{\kappa}'= \Omega'^\dagger {\mathbf{g}}$.

{One best fit is equal to the other one and viceversa. To see this, assume to first compute the coefficients $\bm{\kappa}'$. 
%that provide the best fit \eqref{kprime}. 
To this set of coefficients corresponds a set of coefficients $\bm{\kappa}$, which can be obtained by solving Eq.\ \eqref{NN}.}

\begin{figure}[H] 
\centering
    \includegraphics[width=.45\textwidth]{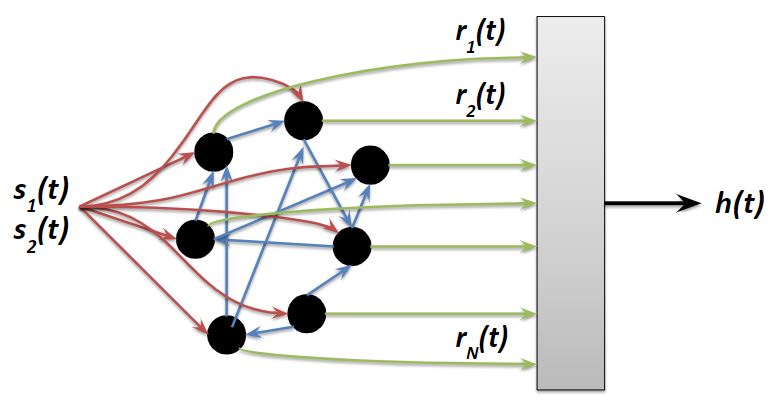}
    \includegraphics[width=.45\textwidth]{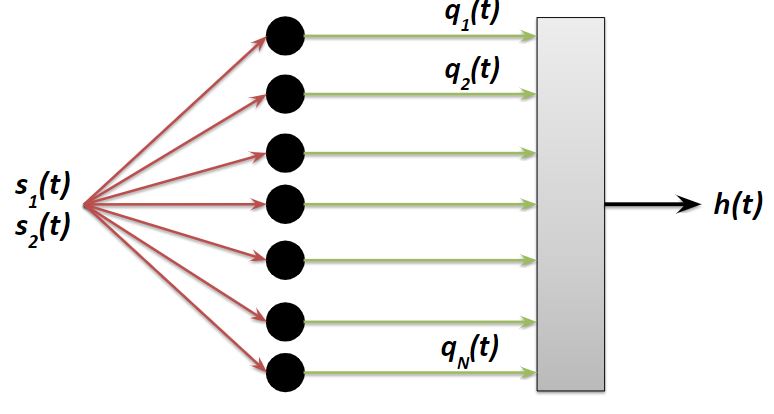}
    \caption{On the left, we have a reservoir computer with a network of connected nodes and on the right a modal decomposition of the reservoir dynamics, where each \emph{mode} is described by a \emph{node}.} \label{key}
\end{figure}

An illustration of the reservoir computer and of its modal decomposition is shown in Fig.\ \ref{key}.
Figure 3 is a plot of the individual modes for the case of the Lorenz system as the parameter $\lambda$ is varied.

There are several advantages of the modal decomposition. One is that in case of instability of one or a few modes, these can be `removed' without the instability affecting the remaining modes. Another advantage is the possibility to individually manipulate each $q_i(t)$ before it is used to generate the best fit to the training signal. One simple such manipulation that we will study in what follows is application of a time shift $q_i(t) \rightarrow q_i(t+\tau_i)$. As we will see, this simple modification will lead to dramatic reductions in the training error.

\begin{figure}[H] \label{mode}
\centering
    \includegraphics[width=.60\textwidth]{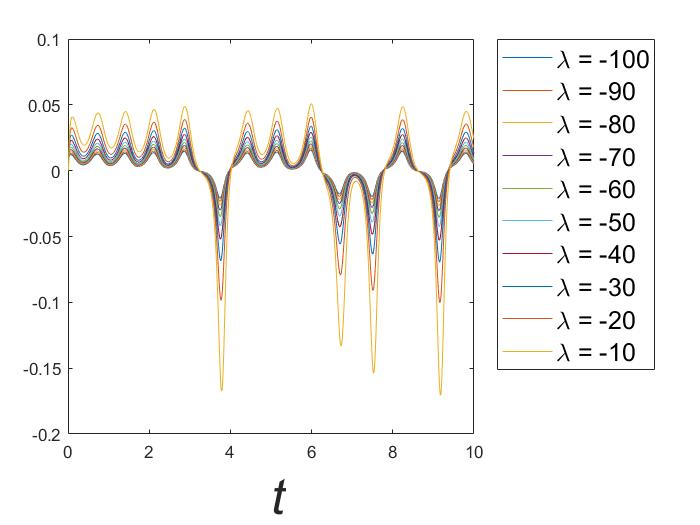}
    \caption{Mode amplitude plot from a continuous time reservoir \eqref{individual node reservior}. The input signal is the $x_L$ state from the Lorenz system. We set $\alpha=-15$ and $\epsilon=0.5$. }
\end{figure}

Our analysis shows that the eigenvalues $\lambda_i$ differentiate individual modes.
We now consider a parametric approach in which the eigenvalues $\lambda_i$ are treated as parameters of Eq. \eqref{individual node reservior}. We first choose an interval $[a,b]$ and then select the $\lambda_i \in [a,b]$. A trivial choice is to pick the eigenvalues to be uniformly randomly distributed in the interval. 

\begin{figure}[H]
\label{robust}
\centering
    \includegraphics[width=.45\textwidth]{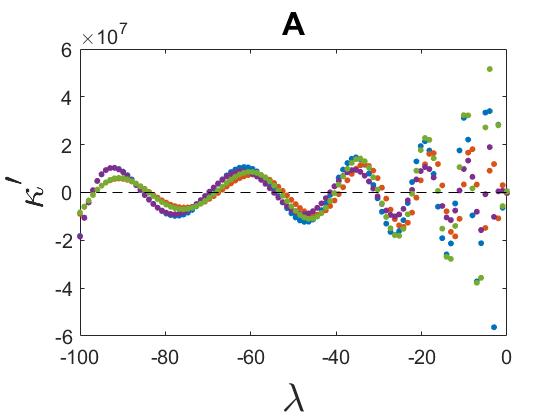}
    \includegraphics[width=.45\textwidth]{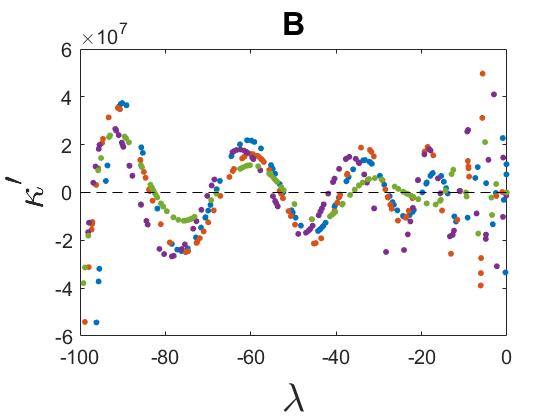}
    \caption{Lorenz system. Continuous time RC, $N=100$ nodes. We use $x_L(t)$ as the input signal and  $z_L(t)$ as the training signal. Linearly spaced eigenvalues are used in plot A and uniformly randomly  distributed eigenvalues are used in plot B. For both plots, eigenvalues are from the interval $[-10^2, -10^{-2}]$.}
\end{figure}

We are now interested in how the coefficients $\kappa'$ (i.e., the mode weights) depend on the particular choice of the eigenvalues $\lambda$. We observe that typically the curve $\kappa'(\lambda)$ is robust to (i) the particular choice of $N$, (ii) the particular sampling of the eigenvalues from the interval $[a,b]$, and (iii) the particular choice of the three time series $s_1(t), s_2(t), g(t)$ from a given (the same) attractor. Property (iii)  indicates robustness of the vector $\pmb \kappa'$ with respect to variations in the initial condition, which predicts a low testing error (see Supplementary Information.)
%As a first experiment, we choose the eigenvalues $\lambda_i$ to be linearly spaced in the interval $[10^{-2},10^2]$ and set all the coefficients $c_i$ to one, we vary the number of nodes to 20, 50, 100, and 200.
As an example of this, in Fig.\ 4 \ref{robust} we plot the resulting $\kappa'$ vs the eigenvalues $\lambda$, for the case of a continuous-time reservoir applied to the Lorenz chaotic system. In plot (A) we used linearly spaced eigenvalues and in plot (B) we use randomly spaced eigenvalues from a uniform distribution. In both plots we set $N=100$ nodes and use the interval $[-10^2, -10^{-2}]$. Different curves in the same plot are for several choices of the initial conditions on the Lorenz attractor. %Each figure is for a specific task, Fig.\ \ref{Lorenz KvsL scatter} is for the Lorenz task, Fig.\ \ref{Hindmash-Rose KvsL scatter} is for the Hindmash-Rose task, and Fig.\ \ref{Thomas KvsL scatter} is for the Thomas task. 
In order to ensure that the time traces $s_1(t), s_2(t), g(t)$ are from the attractor, we take the last point from the Lorenz attractor for the previous iteration as the initial point for the new iteration. In all cases, we see that certain eigenvalues are associated with larger $\kappa'$ values in modulus (both positive and negative.) Other eigenvalues instead have associated $\kappa'$ close to zero, indicating that these  do not play a significant role in the mapping between the input signals and the output (training) signal. We also see that the plots in Fig.\ 4 are consistent over different iterations of the same task, indicating that the \textit{preference} for certain eigenvalues is robust with respect to the particular choice of the input and training time series from the same attractor. The figure also shows that the functional relationship between $\lambda$ and $\kappa'$ is robust with respect to variations in the number of nodes $N$. We thus envision an advantage of picking the eigenvalues $\lambda$'s about the maxima and minima of the $\kappa'$ vs. $\lambda$ plot, which provides the motivation for the optimization study presented next.

\begin{figure}[H] 
\centering
    \includegraphics[width=\textwidth]{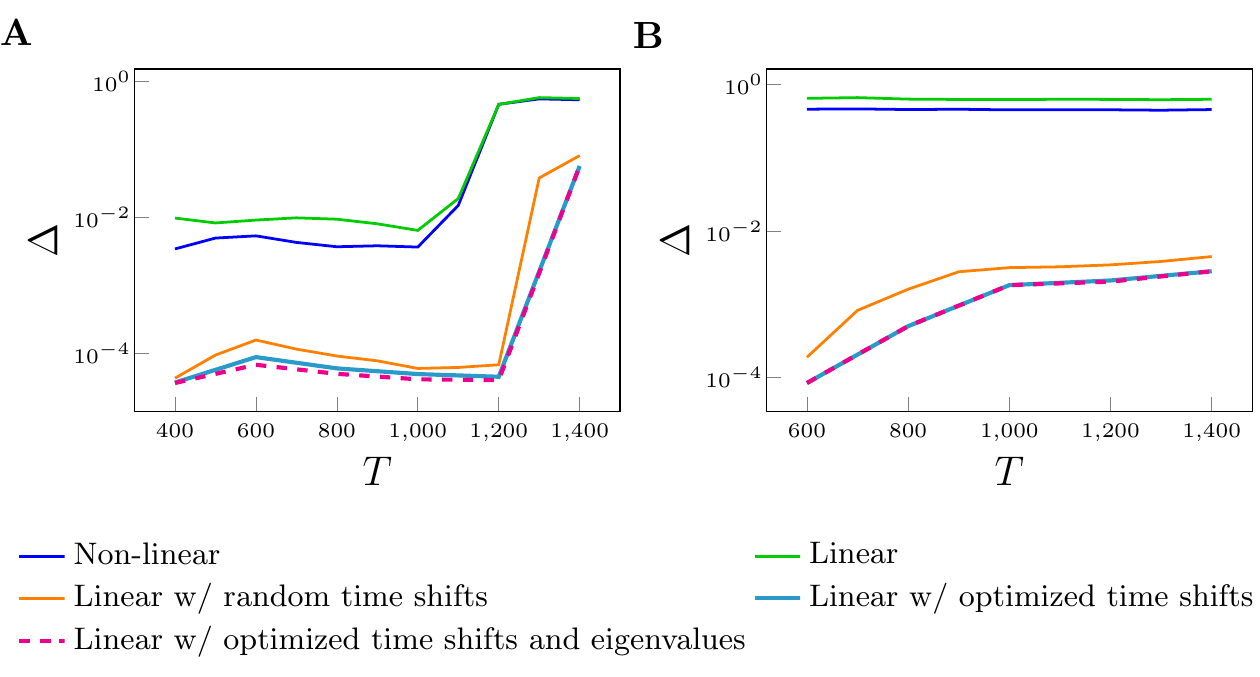}
    \caption{We plot the training error $\Delta$ vs the length of the time series $T$. (A) The HR system. We set $\alpha = -15$, $\epsilon = 20$. For the linear RC $p_2 = p_3= 0$, $c_i =1$ and for the non-linear RC $p_2 = -5$,$p_3= -4$, $c_i =400$.  (B) The Lorenz system. We set $\alpha = -15$, $\epsilon = 0.5$. For the linear RC $p_2 = p_3= 0$, $c_i =1$ and for the non-linear RC $p_2 = -5$, $p_3= -4$, $c_i =30$. In each plot we compare the following cases: nonlinear, linear, linear with application of random time shifts to the individual modes, linear with optimized time shifts of the individual modes, and linear with optimized time shifts and eigenvalues of the individual modes. The time shifts ($\tau_i$) are taken from the interval $[-0.5, 0.5]$ and the eigenvalues ($\lambda_i$) are taken from the interval $[-100, 0]$.} \label{final}
\end{figure}

We now consider a comparison of linear and nonlinear reservoirs, described by the following equations
 \cite{carroll2019network},
\begin{equation} \label{pol}
    \dot q_i(t) =  p_1(t) q_i(t)+ p_2 q_i^2(t) + p_3 q_i^3(t) + \lambda_i q_i(t) + c_i s_1(t), \quad i=1,...,N.
\end{equation}
where $p_2=-5$ and $p_3=-4$ were chosen so as to yield a low training error \cite{doi:10.1063/1.5123733}.  {Note that Eq.\ \eqref{pol} has the same parametric form in $\lambda_i$ as \eqref{individual node reservior} for direct comparison.} When the parameters $c_i$ are small, Eq.\ \eqref{pol} is well approximated by the linear reservoir \eqref{individual node reservior} ($p_2=p_3=0$.)
In Fig.\ \ref{final} we plot the training error $\Delta$ vs the length of the time series $T$. Plot A is for the HR system and plot B is for the Lorenz system. In each plot, we
compare the following cases: nonlinear (Eq.\ \eqref{pol}), linear (Eq.\ \eqref{individual node reservior}), linear  with application of random time shifts to the individual modes, linear with optimized time shifts of the individual modes, and linear with optimized time shifts and eigenvalues of the individual modes. Optimization of the time-shifts and of the eigenvalues was obtained via simulated annealing (see Supplementary Information.)  From Fig.\ \ref{final}, we see that the nonlinear reservoir performs better than the linear one, which is expected \cite{boyd1985fading,bollt2020explaining}. However, this relation is inverted when manipulations of the individual modes of the linear reservoir are introduced, either in terms of the eigenvalues or the time shifts, for which the training error is much lower (also in the case of $\epsilon=0$, see Supplementary Information.) A considerable reduction in the training error is observed even when the time shifts applied to the individual modes are randomly chosen.   In the Supplementary Information we show that a linear reservoir with random time-shifts has both lower training error and testing error than a nonlinear reservoir.

  In this paper we have studied a special class of reservoir computers for which a modal decomposition is possible. This is equivalent to replacing the reservoir network with a set of uncoupled nodes, each one corresponding to a `mode’ of the reservoir. We then have shown that the training error for the two reservoir computers (coupled network and uncoupled nodes) is the same. We build on this result and show that the modes can be manipulated to significantly decrease the overall training error. For example, the simple application of time shifts to the individual modes is found to be highly beneficial; namely, a linear reservoir formed of uncoupled nodes with application of random time shifts to the nodes' outputs, is highly competitive against a nonlinear reservoir. As shown in Fig.\ 5, sometimes, the improvement is by orders of magnitude. A considerable reduction in the training error was observed even when the time shifts applied to the individual modes were randomly chosen. It is worth noting that the ability to either temporally delaying or advancing the individual modes is limited to the uncoupled nodes configuration (right panel of Fig.\ 2), as in the coupled network configuration (left panel of Fig.\ 2), the reservoir states are linear combinations of the modes.

\bibliographystyle{unsrt}
	\bibliography{reservoir,biblio}

\begin{thebibliography}{10}

\bibitem{jaeger2001echo}
Herbert Jaeger.
\newblock The “echo state” approach to analysing and training recurrent
  neural networks-with an erratum note.
\newblock {\em Bonn, Germany: German National Research Center for Information
  Technology GMD Technical Report}, 148(34):13, 2001.

\bibitem{schrauwen2007overview}
Benjamin Schrauwen, David Verstraeten, and Jan Van~Campenhout.
\newblock An overview of reservoir computing: theory, applications and
  implementations.
\newblock In {\em Proceedings of the 15th european symposium on artificial
  neural networks. p. 471-482 2007}, pages 471--482, 2007.

\bibitem{natschlager2002liquid}
Thomas Natschl{\"a}ger, Wolfgang Maass, and Henry Markram.
\newblock The" liquid computer": A novel strategy for real-time computing on
  time series.
\newblock {\em Special issue on Foundations of Information Processing of
  TELEMATIK}, 8(ARTICLE):39--43, 2002.

\bibitem{maass2002real}
Wolfgang Maass, Thomas Natschl{\"a}ger, and Henry Markram.
\newblock Real-time computing without stable states: A new framework for neural
  computation based on perturbations.
\newblock {\em Neural computation}, 14(11):2531--2560, 2002.

\bibitem{martinenghi2012photonic}
Romain Martinenghi, Sergei Rybalko, Maxime Jacquot, Yanne~K Chembo, and Laurent
  Larger.
\newblock Photonic nonlinear transient computing with multiple-delay wavelength
  dynamics.
\newblock {\em Physical review letters}, 108(24):244101, 2012.

\bibitem{brunner2013parallel}
Daniel Brunner, Miguel~C Soriano, Claudio~R Mirasso, and Ingo Fischer.
\newblock Parallel photonic information processing at gigabyte per second data
  rates using transient states.
\newblock {\em Nature communications}, 4:1364, 2013.

\bibitem{nakajima2015information}
Kohei Nakajima, Helmut Hauser, Tao Li, and Rolf Pfeifer.
\newblock Information processing via physical soft body.
\newblock {\em Scientific reports}, 5:10487, 2015.

\bibitem{hermans2015photonic}
Michiel Hermans, Miguel~C Soriano, Joni Dambre, Peter Bienstman, and Ingo
  Fischer.
\newblock Photonic delay systems as machine learning implementations.
\newblock {\em Journal of Machine Learning Research}, 2015.

\bibitem{vinckier2015high}
Quentin Vinckier, Fran{\c{c}}ois Duport, Anteo Smerieri, Kristof Vandoorne,
  Peter Bienstman, Marc Haelterman, and Serge Massar.
\newblock High-performance photonic reservoir computer based on a coherently
  driven passive cavity.
\newblock {\em Optica}, 2(5):438--446, 2015.

\bibitem{duport2016fully}
Fran{\c{c}}ois Duport, Anteo Smerieri, Akram Akrout, Marc Haelterman, and Serge
  Massar.
\newblock Fully analogue photonic reservoir computer.
\newblock {\em Scientific reports}, 6:22381, 2016.

\bibitem{larger2017high}
Laurent Larger, Antonio Bayl{\'o}n-Fuentes, Romain Martinenghi, Vladimir~S
  Udaltsov, Yanne~K Chembo, and Maxime Jacquot.
\newblock High-speed photonic reservoir computing using a time-delay-based
  architecture: Million words per second classification.
\newblock {\em Physical Review X}, 7(1):011015, 2017.

\bibitem{suykens2012artificial}
Johan~AK Suykens, Joos~PL Vandewalle, and Bart~L de~Moor.
\newblock {\em Artificial neural networks for modelling and control of
  non-linear systems}.
\newblock Springer Science \& Business Media, 2012.

\bibitem{crutchfield2010introduction}
James~P Crutchfield, William~L Ditto, and Sudeshna Sinha.
\newblock Introduction to focus issue: intrinsic and designed computation:
  information processing in dynamical systems—beyond the digital hegemony,
  2010.

\bibitem{rodriguez2001simple}
Paul Rodriguez.
\newblock Simple recurrent networks learn context-free and context-sensitive
  languages by counting.
\newblock {\em Neural computation}, 13(9):2093--2118, 2001.

\bibitem{gers2001lstm}
Felix~A Gers and E~Schmidhuber.
\newblock Lstm recurrent networks learn simple context-free and
  context-sensitive languages.
\newblock {\em IEEE Transactions on Neural Networks}, 12(6):1333--1340, 2001.

\bibitem{lu2018attractor}
Zhixin Lu, Brian~R Hunt, and Edward Ott.
\newblock Attractor reconstruction by machine learning.
\newblock {\em Chaos: An Interdisciplinary Journal of Nonlinear Science},
  28(6):061104, 2018.

\bibitem{zimmermann2018observing}
Roland~S Zimmermann and Ulrich Parlitz.
\newblock Observing spatio-temporal dynamics of excitable media using reservoir
  computing.
\newblock {\em Chaos: An Interdisciplinary Journal of Nonlinear Science},
  28(4):043118, 2018.

\bibitem{antonik2018using}
Piotr Antonik, Marvyn Gulina, Ja{\"e}l Pauwels, and Serge Massar.
\newblock Using a reservoir computer to learn chaotic attractors, with
  applications to chaos synchronization and cryptography.
\newblock {\em Physical Review E}, 98(1):012215, 2018.

\bibitem{jaeger2004harnessing}
Herbert Jaeger and Harald Haas.
\newblock Harnessing nonlinearity: Predicting chaotic systems and saving energy
  in wireless communication.
\newblock {\em science}, 304(5667):78--80, 2004.

\bibitem{pathak2017using}
Jaideep Pathak, Zhixin Lu, Brian~R Hunt, Michelle Girvan, and Edward Ott.
\newblock Using machine learning to replicate chaotic attractors and calculate
  lyapunov exponents from data.
\newblock {\em Chaos: An Interdisciplinary Journal of Nonlinear Science},
  27(12):121102, 2017.

\bibitem{pathak2018model}
Jaideep Pathak, Brian Hunt, Michelle Girvan, Zhixin Lu, and Edward Ott.
\newblock Model-free prediction of large spatiotemporally chaotic systems from
  data: A reservoir computing approach.
\newblock {\em Physical review letters}, 120(2):024102, 2018.

\bibitem{jalalvand2018application}
Azarakhsh Jalalvand, Kris Demuynck, Wesley De~Neve, and Jean-Pierre Martens.
\newblock On the application of reservoir computing networks for noisy image
  recognition.
\newblock {\em Neurocomputing}, 277:237--248, 2018.

\bibitem{graves2004biologically}
Alex Graves, Douglas Eck, Nicole Beringer, and Juergen Schmidhuber.
\newblock Biologically plausible speech recognition with lstm neural nets.
\newblock In {\em International Workshop on Biologically Inspired Approaches to
  Advanced Information Technology}, pages 127--136. Springer, 2004.

\bibitem{robinson1994application}
Tony Robinson.
\newblock An application of recurrent nets to phone probability estimation.
\newblock {\em IEEE transactions on Neural Networks}, 5(2), 1994.

\bibitem{lukovsevivcius2012reservoir}
Mantas Luko{\v{s}}evi{\v{c}}ius, Herbert Jaeger, and Benjamin Schrauwen.
\newblock Reservoir computing trends.
\newblock {\em KI-K{\"u}nstliche Intelligenz}, 26(4):365--371, 2012.

\bibitem{carroll2019network}
Thomas~L Carroll and Louis~M Pecora.
\newblock Network structure effects in reservoir computers.
\newblock {\em Chaos: An Interdisciplinary Journal of Nonlinear Science},
  29(8):083130, 2019.

\bibitem{boyd1985fading}
Stephen Boyd and Leon Chua.
\newblock Fading memory and the problem of approximating nonlinear operators
  with volterra series.
\newblock {\em IEEE Transactions on circuits and systems}, 32(11):1150--1161,
  1985.

\bibitem{bollt2020explaining}
Erik Bollt.
\newblock On explaining the surprising success of reservoir computing
  forecaster of chaos? the universal machine learning dynamical system with
  contrasts to var and dmd.
\newblock {\em arXiv preprint arXiv:2008.06530}, 2020.

\bibitem{schmid2010dynamic}
Peter~J Schmid.
\newblock Dynamic mode decomposition of numerical and experimental data.
\newblock {\em Journal of fluid mechanics}, 656:5--28, 2010.

\bibitem{grigoryeva2018universal}
Lyudmila Grigoryeva and Juan-Pablo Ortega.
\newblock Universal discrete-time reservoir computers with stochastic inputs
  and linear readouts using non-homogeneous state-affine systems.
\newblock {\em The Journal of Machine Learning Research}, 19(1):892--931, 2018.

\bibitem{doi:10.1063/1.5123733}
Afroza Shirin, Isaac~S. Klickstein, and Francesco Sorrentino.
\newblock Stability analysis of reservoir computers dynamics via lyapunov
  functions.
\newblock {\em Chaos: An Interdisciplinary Journal of Nonlinear Science},
  29(10):103147, 2019.

\end{thebibliography}

\end{document}